\documentclass{desyproc}

\usepackage{graphicx}
\usepackage{epstopdf}
\usepackage{pdfpages}

\begin{document}
\title{GPGPU for track finding in High Energy Physics}

\author{{\slshape L Rinaldi$^1$, M Belgiovine$^1$, R Di Sipio$^1$, A Gabrielli$^1$, M Negrini$^2$, F Semeria$^2$, A Sidoti$^2$, S A Tupputi$^3$, M Villa$^1$}\\[1ex]
$^1$Bologna University and INFN, via Irnerio 46, 40127 Bologna, Italy\\
$^2$INFN-Bologna,v.le Berti Pichat 6/2, 40127 Bologna, Italy\\
$^3$INFN-CNAF,v.le Berti Pichat 6/2, 40127 Bologna, Italy}

\contribID{37}

\confID{} 
\desyproc{DESY-PROC-2014-05}
\acronym{GPUHEP2014} 
\doi 

\maketitle

\begin{abstract}
The LHC experiments are designed to detect large amount of physics events produced with a very high rate. Considering the future upgrades, the data acquisition rate will become even higher and new computing paradigms must be adopted for fast data-processing: General Purpose Graphics Processing Units (GPGPU) is a novel approach based on massive parallel computing. The intense computation power provided by Graphics Processing Units  (GPU) is expected to reduce the computation time and to speed-up the low-latency applications used for fast decision taking. In particular, this approach could be hence used for high-level triggering in very complex environments, like the typical inner tracking systems of the multi-purpose experiments at  LHC, where a large number of charged particle tracks  will be produced with the luminosity upgrade.
In this article we discuss a track pattern recognition algorithm based on the Hough Transform, where a parallel approach is expected to reduce dramatically the execution time.
\end{abstract}

\section{Introduction}

Modern High Energy Physics (HEP) experiments are designed to detect large amount of data with very high rate. In addition to that weak signatures of new physics must be searched in complex background condition. In order to reach these achievements, new computing paradigms must be adopted. A novel approach is based on the use of high parallel computing devices, like  Graphics Processing Units (GPU), which delivers such high performance solutions to be used in HEP. In particular, a massive parallel computation based on General Purpose Graphics Processing Units (GPGPU)~\cite{nvidia} could dramatically speed up the algorithms for charged particle tracking and fitting, allowing their use for fast decision taking and triggering. 
In this paper we describe a tracking recognition algorithm based on the Hough Transform~\cite{hough:paper,hough:hep1,hough:hep2} and its implementation on Graphics Processing Units (GPU).

\section{Tracking with the Hough Transform}

The Hough Transform (HT) is a pattern recognition technique for features extraction in image processing, and in our case we will use a HT based algorithm to extract the tracks parameters from the hits left by charged particles in the detector. A preliminary result on this study has been already presented in~\cite{tipp}. Our model is based on a cylindrical multi-layer silicon detector installed around the interaction point of a particle collider, with the detector axis on the beam-line. 
The algorithm works in two serial steps. In the first part, for each hit having coordinates $(x_H,y_H,z_H)$ the algorithm computes all the circles in the $x-y$ transverse plane passing through that hit and the interaction point, where the circle equation is $x^2+y^2-2Ax-2By=0$, and $A$ and $B$ are the two parameters corresponding to the coordinates of the circle centre. The circle detection is performed taking into account also the longitudinal ($\theta$) and polar ($\phi$) angles. For all the $\theta$, $\phi$, $A$, $B$, satisfying the circle equation associated to a given hit, the corresponding $M_H(A,B,\theta,\phi)$ Hough Matrix (or Vote Matrix) elements are incremented by one. After computing all the hits, all the $M_H$ elements above a given threshold would correspond to real tracks. Thus, the second step is a local maxima search among the $M_H$ elements.

In our test, we used a dataset of 100 simulated events ($pp$ collisions at LHC energy, Minimum Bias sample with tracks having transverse momentum $p_T>500$ MeV), each event containing up to 5000 particle hits on a cylindrical 12-layer silicon detector centred on the nominal collision point. The four hyper-dimensions of the Hough space have been binned in $4 \times 16 \times 1024 \times 1024$ along the corresponding $A,B,\theta,\phi$ parameters.

\begin{figure}[h]
\centerline{\includegraphics[width=.8\textwidth]{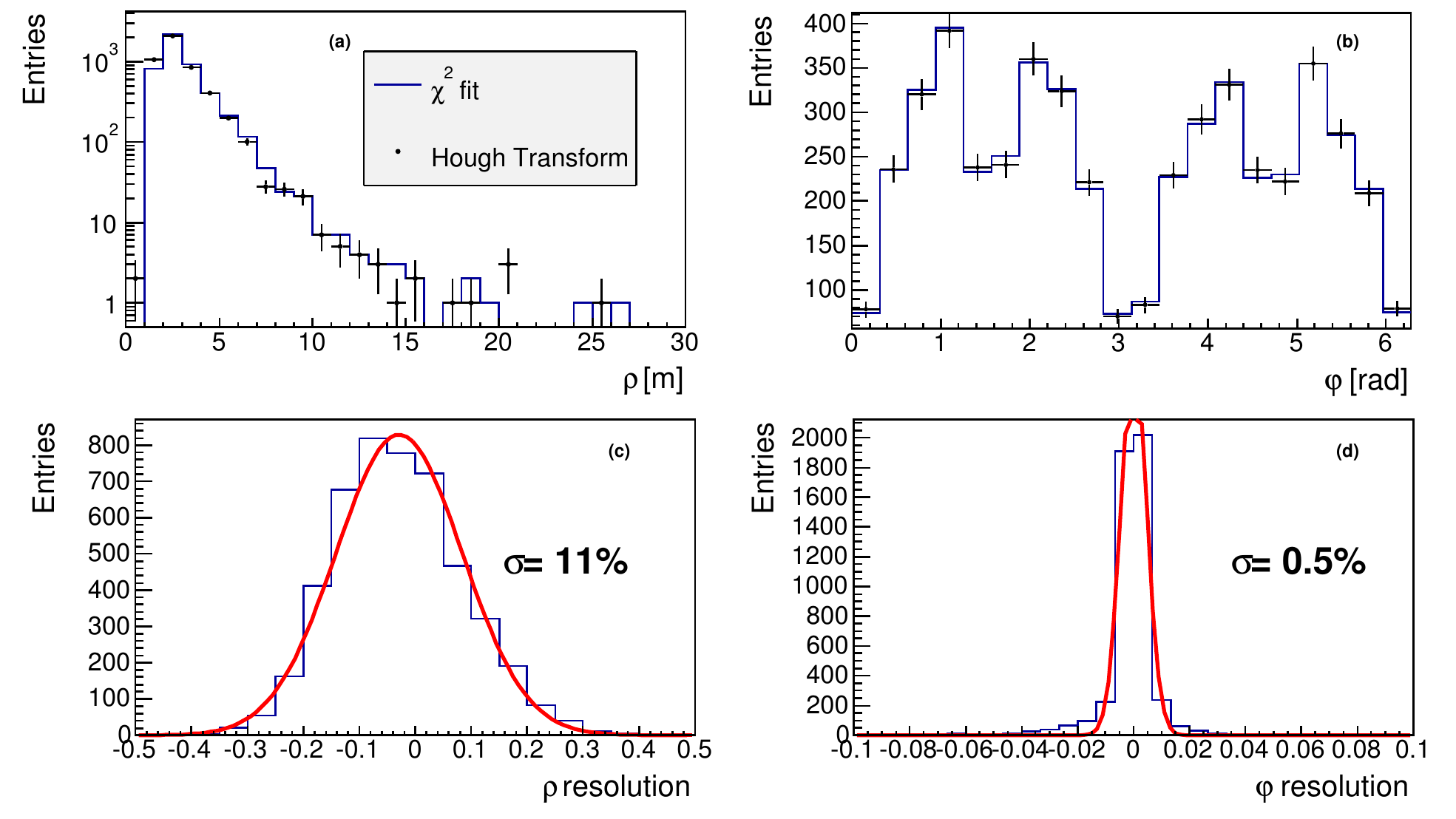}}
\caption{Hough Transform algorithm compared to $\chi^2$ fit. (a) $\rho$ distribution; (b) $\varphi$ distribution; (c) $\rho$ resolution; (d) $\varphi$ resolution. }\label{hough:perf}
\end{figure}

The algorithm performance compared to a $\chi^2$ fit method is shown in Fig.~\ref{hough:perf}: the $\rho=\sqrt{A^2+B^2}$ and $\varphi=\tan^{-1}(B/A)$ are shown together with the corresponding resolutions.

\section{SINGLE-GPU implementation}

\begin{figure}[hb]
\centerline{\includegraphics[width=.6\textwidth]{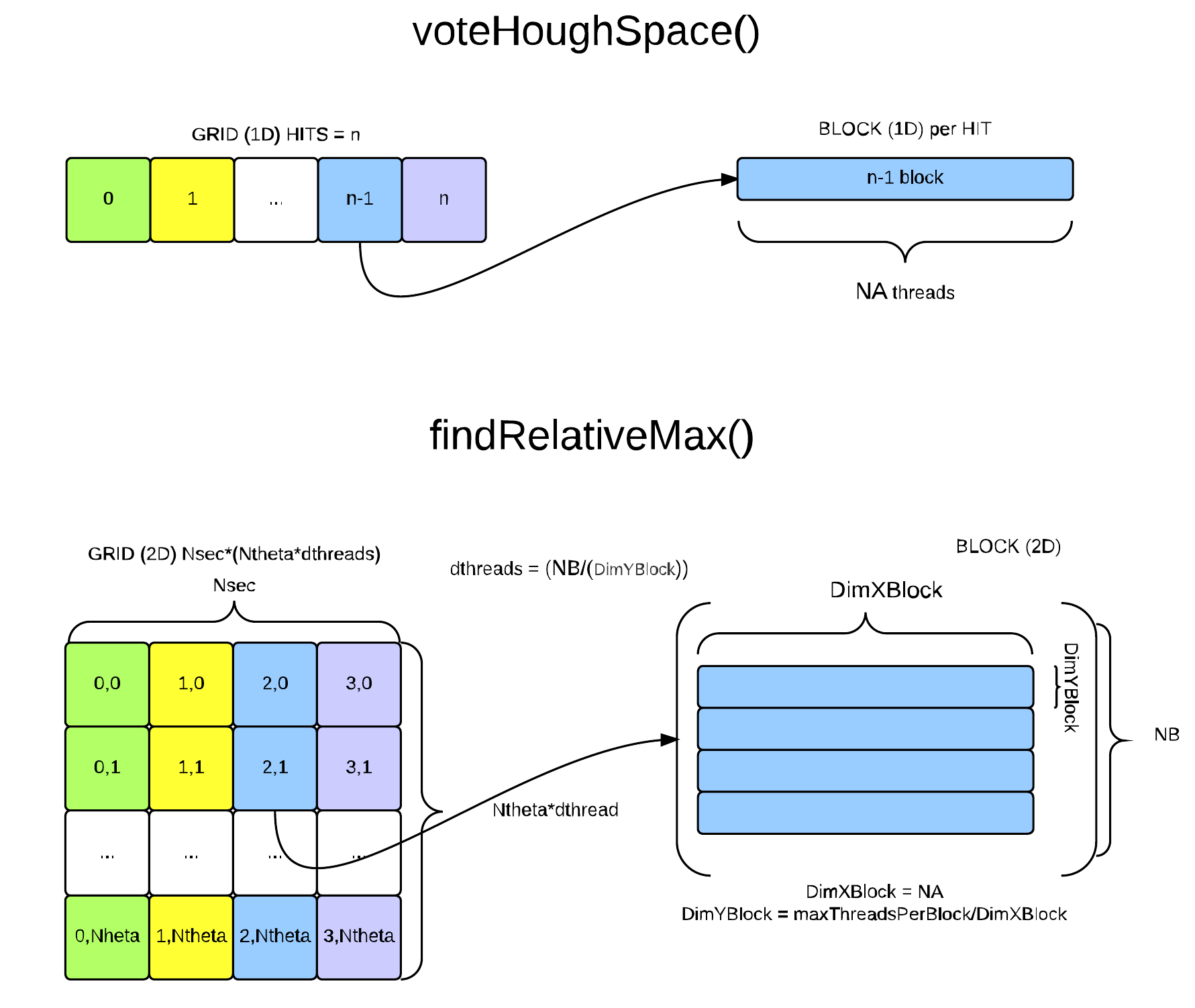}}
\caption{GPGPU implementation schema of the two Hough Transform algorithm kernels.}\label{GPGPU:schema}
\end{figure}

The HT tracking algorithm has been implemented in GPGPU splitting the code in two kernels, for Hough Matrix filling and searching local maxima on it. Implementation has been performed both in CUDA~\cite{nvidia} and OpenCL~\cite{opencl}. GPGPU implementation schema is shown in Fig.~\ref{GPGPU:schema}.

Concerning the CUDA implementation, for the $M_H$ filling kernel, we set a 1-D grid over all the hits, the grid size being equal to the number of hits of the event. Fixed the ($\theta,\phi$) values, a thread-block has been assigned to the $A$ values, and for each $A$, the corresponding $B$ is evaluated. The $M_H(A,B,\theta,\phi)$ matrix element is then incremented by a unity with an {\tt atomicAdd} operation. The $M_H$ initialisation is done once at first iteration with {\tt cudaMallocHost} (pinned memory) and initialised on device with {\tt cudaMemset}. 
In the second kernel, the local maxima search is carried out using a 2-D grid over the $\theta,\phi$ parameters, the grid dimension being the product of all the parameters number over the maximum number of threads per block ($N_{\phi} \times N_{\theta} \times N_A \times N_B$)/{\tt maxThreadsPerBlock}, and 2-D threadblocks, with {\tt dimXBlock}=$N_A$ and {\tt dimYBlock=MaxThreadPerBlock}/$N_A$. Each thread compares one $M_H(A,B,\theta,\phi)$ element to its neighbours and, if the biggest, it is stored in the GPU shared memory and eventually transferred back. With such big arrays the actual challenge lies in optimizing array allocation and access and indeed for
this kernel a significant speed up has been achieved by tuning matrix access in a coalesced fashion, thus allowing to gain a crucial computational speed-up.
The OpenCL implementation has been done using a similar structure used for CUDA. Since in OpenCL there is no direct pinning memory, a device buffer is mapped to an already existing $memallocated$ host buffer ({\tt clEnqueueMapBuffer}) and dedicated kernels are used for matrices initialisation in the device memory. The memory host-to-device buffer allocation is performed concurrently and asynchronously, saving overall transferring time.

\subsection{SINGLE-GPU results}

\begin{footnotesize}
\begin{table}[h]
\centerline{\begin{tabular}{ | l | c | c | c |}
\hline
Device & NVIDIA & NVIDIA & NVIDIA \\
specification & GeForce GTX770 & Tesla K20m & Tesla K40m \\
\hline
Performance (Gflops) & 3213 & 3542 & 4291 \\
Mem. Bandwidth (GB/s) & 224.2 & 208 & 288 \\
Bus Connection & PCIe3 & PCIe3 & PCIe3 \\
Mem. Size (MB) & 2048 & 5120 & 12228 \\
Number of Cores & 1536 & 2496 & 2880 \\
Clock Speed (MHz) & 1046 & 706 & 1502 \\
\hline
\end{tabular}}
\caption{Computing resources setup.}
\label{tab:gpus}
\end{table}
\end{footnotesize}

The test has been performed using the NVIDIA~\cite{nvidia} GPU boards listed in table~\ref{tab:gpus}. The GTX770 board is mounted locally on a desktop PC, the Tesla K20 and K40 are installed in the INFN-CNAF HPC cluster.

\begin{figure}[h]
\centerline{\includegraphics[width=1.05\textwidth]{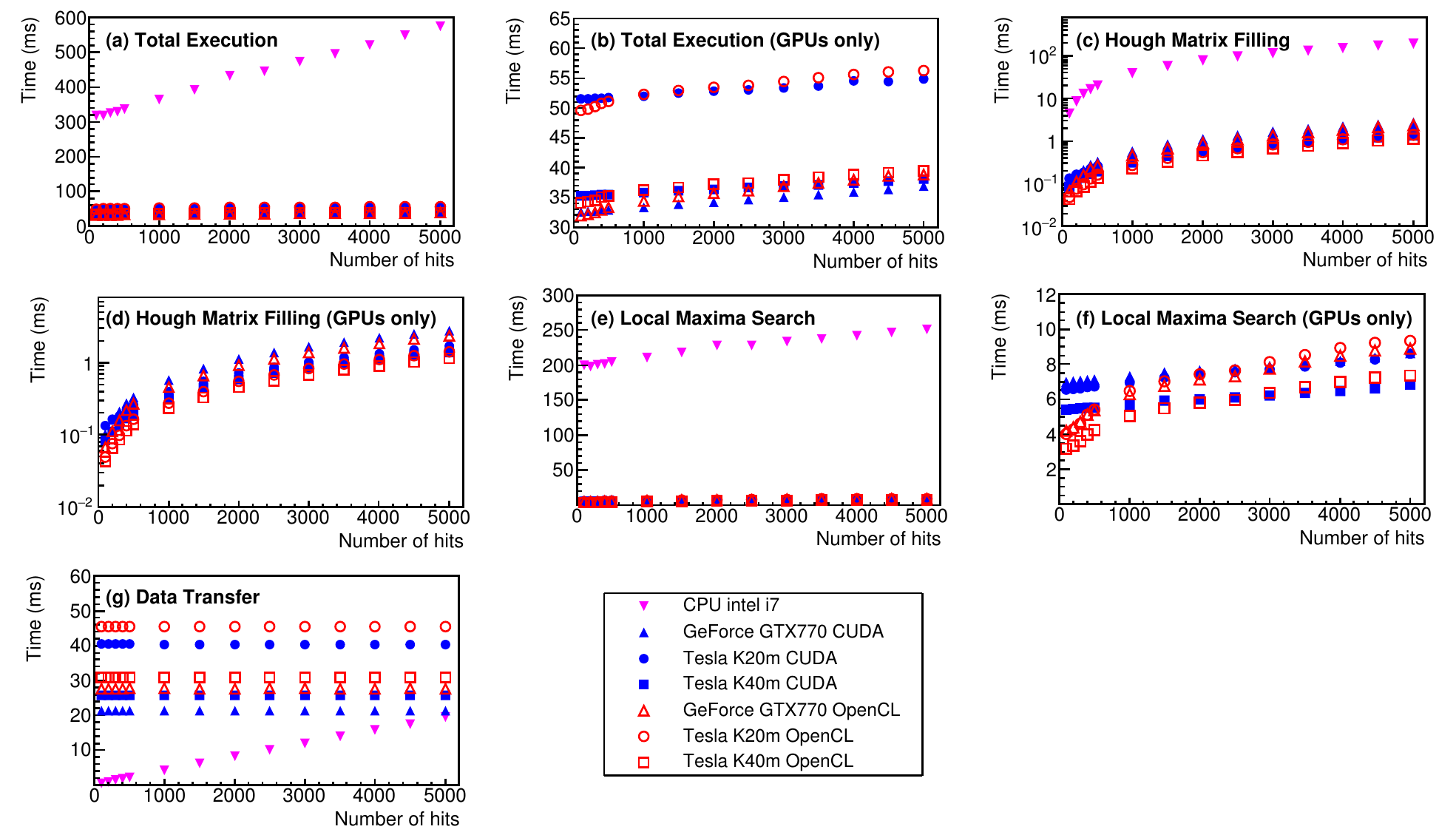}}
\caption{Execution timing as a function of the number of analysed hits. (a) Total execution time for all devices; (b) Total execution time for GPU devices only; (c) $M_H$ filling time for all devices; (d) $M_H$ filling timing for GPU devices only; (e) local maxima search timing for all devices; (f) local maxima search timing for GPU devices only; (g) device-to-host transfer time (GPUS) and I/O time (CPU). }\label{allgpu}
\end{figure}

The measurement of the execution time of all the algorithm components has been carried out as a function of the number of hits to be processed, and averaging the results over 100 independent runs. The result of the test is summarised in Fig.~\ref{allgpu}.
The total execution time comparison between GPUs and CPU is shown in Fig.~\ref{allgpu}a, while in Fig.~\ref{allgpu}b the details about the execution on different GPUs are shown. The GPU execution is up to 15 times faster with respect to the CPU implementation, and the best result is obtained for the CUDA algorithm version on the GTX770 device. The GPUs timing are less dependent on the number of the hits with respect to CPU timing.

The kernels execution on GPUs is even faster with respect to CPU timing, with two orders of magnitude GPU-CPU speed up, as shown in Figs.~\ref{allgpu}c and \ref{allgpu}e. When comparing the kernel execution on different GPUs (Figs.~\ref{allgpu}d) and~\ref{allgpu}f), CUDA is observed to perform slightly better than OpenCL. Figure~\ref{allgpu}g shows the GPU-to-CPU data transfer timings for all devices together with the CPU I/O timing, giving a clear idea of the dominant part of the execution time.

\section{MULTI-GPU implementation}

Assuming that the detector model we considered could have multiple readout boards working independently, it is interesting to split the workload on multiple GPUs. We have done this by splitting the transverse plane in four sectors to be processed separately, since the data across sectors are assumed to be read-out independently. 
Hence, a single HT is executed for each sector, assigned to a single GPU, and eventually the results are merged when each GPU finishes its own process. The main advantage is to reduce the load on a single GPU  by using lightweight Hough Matrices and output structures. Only CUDA implementation has been tested, using the same workload schema discussed in Sec. 3, but using four $M_H(A,B,\theta)$, each matrix processing the data of a single $\phi$ sector.
\begin{figure}[h]
\centerline{\includegraphics[width=.8\textwidth]{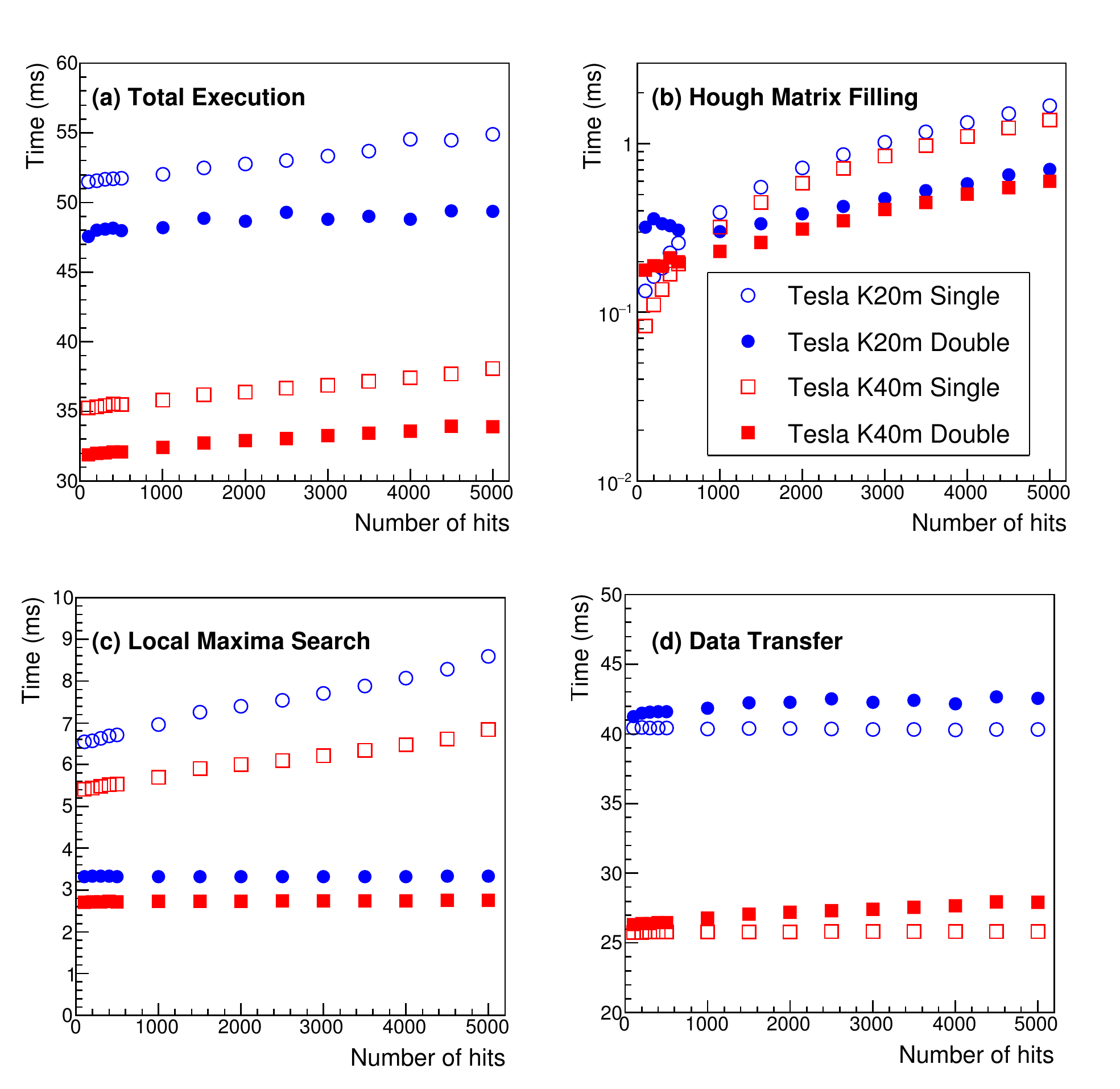}}
\caption{Execution timing as a function of the number of the hits for multi-GPU configuration. (a) Total execution time; (b) $M_H$ filling timing; (c) local maxima search timing; (d) device-to-host transfer time.}\label{multigpu}
\end{figure}

\subsection{MULTI-GPU results}

The multi-GPU results are shown in Fig.~\ref{multigpu}. The test has been carried out in double configuration, separately, with two NVIDIA Tesla K20 and two NVIDIA Tesla K40. The overall execution time is faster with double GPUs in both cases, even if timing does not scale with the number of GPUs. An approximate half timing is instead observed when comparing kernels execution times. On the other hand, the transferring time is almost independent on the number of GPUs, this leading the overall time execution.

\section{Conclusions}

A pattern recognition algorithm based on the Hough Transform has been successfully implemented on CUDA and OpenCL, also using multiple devices. The results presented in this paper show that the employment of GPUs in situations where time is critical for HEP,
like triggering at hadron colliders, can lead to significant and encouraging speed-up. Indeed the problem by itself offers wide room for a parallel approach to computation: this is reflected in the results shown where the speed-up is around 15 times better than what achieved with a normal CPU. There are still many handles for optimising the performance, also taking into account the GPU architecture and board specifications. 
Next steps of this work go towards an interface to actual experimental frameworks, including the management of the experimental data structures and testing with more graphics accelerators and coprocessor.


\begin{footnotesize}

\end{footnotesize}

\end{document}